\documentclass{article}

\usepackage{authblk}
\usepackage[in]{fullpage}
\usepackage{graphicx}
\usepackage{amssymb}

\newcommand{\cerenkov}{Cherenkov }

\begin{document}

\title{Sub-Penning gas mixtures: new possibilities for ton- to kiloton-scale time projection chambers}
\author{Benjamin Monreal}
\author{Luiz de Viveiros}
\author{William Luszczak}
\affil{University of California, Santa Barbara}

\maketitle

\begin{abstract}
  In this work, we present the concept for large low-background experiments in which an unusual gas mixture gas serves as a seamless, high-QE, near-100\%-coverage photodetector for scintillation or \cerenkov photons.    We fill a large time projection chamber with a VUV scintillating gas, plus an unusually small admixture dopant gas with a low ionization threshhold (and a high ionization yield), akin to a highly-underquenched Penning mixture.   Scintillation photons travel far from a primary ionization site before converting into photoionization electrons.   Using standard TPC methods, we can separately count both the primary ionization electrons (which occur along a dense track) and the scintillation-ionization electrons (which will occur over a large spherical region) without the use of PMTs.   The scheme is compatible with very large detectors, in both two-phase and single-phase high pressure configurations.   We discuss how the drift-axis position of an event can be reconstructed, and under what constraints we can expect stable gas gain operations.  We propose some detectors illustrating how this scheme---both in conventional two-phase geometries, as well as in pressurized space in solution-mined salt cavern---makes it possible to safely construct gas time projection chambers of previously-unreachable target masses, capable of studying dark matter, double beta decay, proton decay, and solar neutrinos; more speculative gas mixtures might extend the technique to reactor and geoneutrinos.   \end{abstract}


\section{General principle}
Organic vapors with very low ionization threshholds, like TEA (triethyl amine) TMA (trimethyl amine), and TMAE (terakis(dimethylamino)ethylene) have been used in particle physics in two ways: as gaseous photocathodes (most commonly TMAE behind a UV-transparent window) in large-area \cerenkov ring imaging (RICH) detectors \cite{Vavra:NuclearInstrumentsAndMethodsInPhysicsResearch:1996}, and as Penning mixtures to maximize ionization yield in gas and liquid ionization chambers \cite{Alvarez:JournalOfInstrumentation:2014}.   The concept of a Penning mixture is that, in (for example) pure Xe, an ionizing radiation track will leave behind many excited Xe$^*$ atoms, and many will deexcite or scintillate.  When we add a sufficient density of (for example) TMAE, the Xe$^*$ atoms collisionally or resonantly transfer energy to TMAE molecules and ionize them, because TMAE has a sufficiently low ionization threshhold.  Due to these energy transfers, the gas mixture has a very high ionization yield (and a high Townsend coefficient for gain) but very little scintillation.   Note that the Penning effect relies on a \emph{high} density of dopant molecules in the mixture, since the Penning transfer is short-range.  What happens in a gas with a \emph{low} density of dopant?   In such a mixture, excited atoms can deexcite by free scintillation as usual.   The photons propagate over a long range, but now they photoionize dopant molecules far from the point of origin.    In some cases, this photoionization has a very high quantum efficiency---far higher than any solid-state photocathode.    Therefore, ionizing radiation in a sub-Penning mixture leaves us with two distinct populations of electrons: primary ionization electrons at the site of the primary event, and photoionization electrons (from scintillation photons interacting with the dopant gas) spatially distributed around the primary site.    This is important for three reasons.  First, in a sub-Penning gas, unlike in a Penning gas, we have the opportunity to measure the separate scintillation and ionization signals which have proven useful in so many other experiments.   Second, as discussed below, we expect to be able to reconstruct useful drift-axis information from details of the photoionization cloud.    Third, we do this without PMTs, since all of the deposited energy is in the form of electrons.    Additionally, we can choose gas mixtures where \cerenkov photons, not ionization photons, generate the photoionization signal.

We note some general benefits of this scheme.   (a) We can build highly capable, fiducialized scintillation/\cerenkov detectors with \emph{no photomultiplier tubes}, only ionization drift and gas amplification electronics\footnote{One might still opt for PMTs as part of, e.g., an electroluminescent charge readout; however, there are many non-PMT options for TPC whereas there are very few for direct scintillation light collection.}.   Removing PMTs from such a system has repercussions for cost, radiopurity, pressure tolerance, geometrical and optical constraints, and many other design issues.  (b) By counting both primary and photoconversion electrons, there is the possibility of reaching very high energy resolutions, possibly approaching those of an ideal Penning mixture; this is particularly important for double-beta decay searches.  (c) If the ionization quantum efficiency is high at the scintillation wavelength---and in some cases yields may approach 100\%---then this may be the ``holy grail'' of an inexpensive, seamless, windowless, 100\%-QE photocathode.   

We propose to take advantage of PMT-free design to build gigantic, and plausibly affordable, detectors for a wide range of astroparticle physics experiments.   In particular, we sketch the construction parameters of  (a) a 30~tonne, low-threshhold two-phase xenon TPC for dark matter, (b) a 32~tonne xenon gas TPC for neutrinoless double beta decay, (c) a kilotonne ultrahigh pressure neon TPC for solar neutrino and dark matter physics, and (now with some speculation about gas properties), (d) a 50 kT two-phase (Ar:Xe):TMA \cerenkov-imaging TPC for accelerator neutrinos.   In less detail, we discuss the uncertain possibility of H$_2$ or CH$_4$-rich TPCs for antineutrino physics.   We invite interested readers to consider other applications.

\subsection{Gas mixture options}

TEA and TMAE are organic molecules with unusually low ionization threshholds and are appropriate dopants for Ar and Xe detectors, and potentially CF$_4$.  Their photoionization quantum efficiencies are moderate (20\%) at low pressures but rise at higher pressures and in the liquid state.   For He and Ne, which produce very short-wavelength scintillation, the noble gases themselves (most appropriately xenon) can serve as dopants.   Importantly, xenon should have 100\% photoionization yield, since it has no non-ionizing excited states above 7 eV.   (TEA and TMAE have numerous photodissociation channels that compete with ionization.)    

In Tables \ref{gasprop1} and \ref{gasprop2} we give the absorption lengths and quantum yields of scintillation light in several suggested sub-Penning mixtures.   Note that Xe:TMAE and Ar:TEA can be made in the ambient-pressure liquid state, but for the most part Ne:Xe and He:Xe cannot because the xenon will freeze out.   Ne:H$_2$ is an option for a liquid-state Ne detector.   

\begin{table}
  \begin{tabular}{|p{0.18\textwidth}|p{0.10\textwidth}|p{0.20\textwidth}|p{0.10\textwidth}|p{0.15\textwidth}|}
    \hline
    Gas : dopant & Scintillation wavelength & Photoabsorption cross section & Ionization quantum yield & Partial pressure for 1~m path\\
    \hline
    He :Xe & 76 nm  & $1\times10^{-17}$~cm$^2$ &  100\% & 5 mTorr \\
    Ne :H$_2$ & 76 nm & $5.4\times 10^{-17}$~cm$^2$ & $\sim$ 100\%& 31 mTorr \\
    Ne :Xe & 76 nm & $1\times10^{-17}$~cm$^2$ &  100\% & 5 mTorr \\
    Ar :TMAE & 129 nm &   $9\times 10^{-17}$~cm$^2$  & 50\% & 10 mTorr \\
    Xe :TMAE & 178 nm & $1.5\times 10^{-17}$~cm$^2$ & 35\%  & 20 mTorr \\ 
    H$_2$ :TMAE & \multicolumn{3}{|c|}{ 185 nm (?) Scintillation present but poorly studied } & 20 mTorr\\
    \hline 
   (CH$_4$:CF$_4$):TMAE &  \multicolumn{3}{|c|}{210 nm (?)  Presence of scintillation is speculative} & 20 mTorr \\ 
\hline
\end{tabular}
  \caption{Suggested sub-Penning gas mixtures for scintillation light.  Examples are given based on He, Ne, Ar, Xe, and H$_2$ with appropriate dopants and their properties.  TEA and TMAE cross sections and ionization yields are given based on low-pressure data from  \cite{Holroyd:NuclearInstrumentsAndMethodsA:1987}. Yields are expected to be much higher in high pressure \cite{Nakagawa:NuclearInstrumentsAndMethodsInPhysicsResearch:1993} and cryogenic liquid  \cite{hitachi1997photoionization}  environments, with liquid Xe:TMAE possibly approaching 100\% yield.  At EUV wavelengths (He, Ne), organic vapors are expected to lose ionization efficiency since a number of fragmentation channels become available, so we opt for atomic gases (here Xe, but Hg would also work) which should have essentially 100\% QE.   Hydrogen gas scintillates with 15 photons/MeV in the range 185--210~nm \cite{Dieterle:NuclearInstrumentsAndMethods:1979} and has not been studied deeper in the VUV.  We speculate that CH$_4$:CF$_4$ mixtures may scintillate.} \label{gasprop1}
  \end{table}

\begin{table} 
\begin{tabular}{|p{0.13\textwidth}|p{0.25\textwidth}|p{0.2\textwidth}|p{0.26\textwidth}|}
    \hline       
  Gas : dopant & \cerenkov sensitivity range & Photon yield ($\gamma$/cm) & photoelectron yield (pe/cm) \\
    \hline       
    (Ar:Xe):TMA & $\sim$130--135 nm  & 60 & 20 \\
    CH$_4$:TMAE & $\sim$155-210 nm& 500 & 150  \\ 
   
    \hline       
  \end{tabular}
  \caption{Suggested sub-Penning mixtures for detection of \cerenkov radiation in liquid phase TPCs.  Pure Ar is a bright scintillator at 129~nm.  An Ar:Xe mixture (requiring Xe $\gtrapprox$ 1\%) transfers the scintillation light longwards of the TEA ionization threshhold \cite{Neumeier:Epl:2015}, but leaves an VUV-transparent window around 130--135~nm in which TEA-ionizing \cerenkov photons can propagate.    The VUV absorption properties of liquid CH$_4$ have not been measured, but extrapolation from the gas state suggests there will be a window of transparency for TMAE-ionizing \cerenkov radiation.}  \label{gasprop2}
  \end{table}

\section{TPC design and construction considerations}

For a given gas and dopant, we can choose the dopant concentration such that the mean free path of a scintillation photon takes any desired value $\lambda$.   A pointlike ionization event will result in an electron cloud with a dense primary ionization track, in the center of a spherically symmetric cloud of single electrons, with an exponential radial profile with scale length $\lambda$, from scintillation photoconversion on the TEA.    The choice of $\lambda$ needs to be informed by several issues, including gain stability and z-axis reconstruction.   Gas gain stability is a complex issue in a single-phase gas detector, but potentially much simpler in a two-phase liquid/gas system.

\subsection{Amplification, gating, and gain instabilities} 

In a gas-phase detector, this particular choice of gas appears, at first glance, extraordinarily prone to a UV-feedback instability \cite{Vavra:NuclearInstrumentsAndMethodsInPhysicsResearch:1996}---a gas-gain avalanche is needed to detect electrons, but the avalanche itself emits UV photons, which generate additional primaries; this is sometimes referred to as ``avalanche breeding'', and leads to (at best) afterpulsing or (at worst) a streamer discharge, if not quenched or gated somehow.   In gas RICH applications, this is avoided with non-scintillating gas mixtures.  In our proposed systems, it can be avoided only by careful design of the gas amplification stack.   Stable operations require the use of ``optically blind'' gain structures; we must aggressively block any open line-of-sight between the gas amplification subvolume and the main drift volume.  A double GEM, operating at zero gain, is a good candidate for such a structure.    However, we must also be careful with the gas volume between the blind and the anode; UV photons must have a low probability of photoconverting in this volume.   The constraint is roughly that the average single-electron avalanche should produce $<1$ new photoconversion electrons in any region that leads to further gain.  Consider a gain avalanche of gain $G$, occurring in an small VUV-blinded area of characteristic size $a$.   The gas avalanche produces $\epsilon G$ photons per primary electron, where $\epsilon$ is the electroluminescent quantum yield of the avalanche.  For stability, then, we demand $\lambda/a < \epsilon G$.    This consideration applies separately to each optically-isolated area; a triple-GEM stack could tolerate $G \lessapprox \lambda/\epsilon a$ in {\em each} of the gain holes.    The parameter $\epsilon$, though hard to generalize, is typically $<1$ in high-field avalanches (for example $\epsilon \simeq 0.016$ in Ne:CF$_4$ in a THGEM \cite{Rubin:2013Jinst8P08001:2013}) but can be arbitrarily large in electroluminescence gaps.  Roughly speaking, a charge/current readout needs the system to have ``large enough'' $G$ while a photon/electroluminescent readout requires a large $\epsilon G$; it is $\epsilon G$ that affects the stability of these gases.   

For example, consider a triple-GEM stack with 1 mm spacing, each 99\% opaque to photons generated in the holes one layer down.  We operate the three GEMS at gains of 0, $\sim$10, and $\sim$500 respectively.   The top GEM is only an optical baffle; the middle GEM operates at low gain in order to have a low chance of sending an ionizing photon into the main drift volume (where it would very likely generate a new photoelectron); the bottom GEM produces many photons per avalanche, but the mechanical structures reabsorb them efficiently within $\sim$1~mm.  Therefore, we can avoid discharges in the region near the bottom GEM as long as $a \epsilon G = \lambda > 500$~mm.  Therefore the system can operate stably if $G = 5000/\epsilon$.   $G=5000$ is certainly sufficient for either pad readout or optical readout with TPB-coated SiPMs.  More speculatively, multilayer Kapton circuit etching methods, like those used to produce microbulk Micromegas \cite{Dafni:NuclearInstrumentsAndMethodsInPhysicsResearch:2009}, can in principle make gain structures with 50--100~$\mu$m baffling of gain avalanche scintillation, and these are compatible with shorter photoconversion lengths.    

A separate stability condition involves ``traditional'' UV feedback, caused by VUV electroluminescence photons that liberate electrons from solid surfaces in the detector; in a sub-Penning TPC, these will generally be surfaces in the baffle region.  Since we are operating at low gain, the effect should be of minor importance.    Also, since  the afterpulsing time constant will generally be short compared with the typical time separation between signal electrons.   It is therefore unlikely that small afterpulses would be easily mistaken for photoionization signals.   We note that some analyses, like scintillation pulse shape discrimination (see section \ref{psd}), may require special precautions.   

Many of the detectors proposed below would operate at high pressures.   Conventional micropattern detectors can produce gain only at moderate pressures (1--10 bar) \cite{bondar2002high}.   Simple wire anodes can operate at much higher pressures \cite{Sood:NuclearInstrumentsAndMethodsA:1994} but need more careful baffle design.  Examples of optically-baffled wire anode planes can be found in  \cite{Vavra:NuclearInstrumentsAndMethodsInPhysicsResearch:1996, Aston:NuclearInstrumentsAndMethodsInPhysicsResearch:1989} among others.
 
Avalanche-free, electroluminescent (EL) gain is desirable for obtaining the highest ionization electron counting fidelity.   Since EL systems typically involve VUV emission inside comparatively large gaps---5~mm in NEXT \cite{Collaboration:Jinst7:2012} for example---they may require fairly long photoconversion lengths ($\sim$5~m), but this is compatible with most of the case studies below.

\subsubsection{Dopant partitioning in a two-phase TPC}
The gain is simpler in the case of a two-phase liquid/gas TPC.  The dopant concentration in the {\em liquid} component determines the scintillation photon conversion length relevant to detection/reconstruction.   The dopant concentration in the {\em gas} component determines the scintillation photon conversion length relevant to gas discharges and gain instability.   In every mixture discussed here, though, the dopant will be almost entirely absent from the gas phase, since the dopant molar density is extremely low\footnote{Although direct data is absent, these are all nonpolar molecules and nonpolar solvents, so we expect good adherence to Raoult's Law.} so the dopant gas-phase partial pressure is suppressed by a very large factor below its liquid-phase density. 

Therefore, in a two-phase TPC, gain avalanches occur in an essentially pure, undoped gas, so there is no gain-stability constraint on the size scale $a$ of the optically-clear space around the avalanche.   We still need to optically baffle the gas avalanche UV photons to prevent them from photoconverting in the liquid, but this is a weaker engineering constraint and can be met using a conventional multi-GEM stack.


\subsection{Drift geometry, z-axis reconstruction}\label{sec_drift_axis} Where does the Z-axis information come from?  Since we have absorbed the scintillation light in the gas, it seems like we have disabled the t=0 signal (or S1) that normally provides a z-axis coordinate (or an r-coordinate in a cylindrical detector) for a drifted ionization event.   On one hand, we could reconstruct $z$ by measuring diffusion of the ionization charge \cite{Lewis:NuclearInstrumentsAndMethodsInPhysicsResearch:2015}, but this requires extremely fine-grained readout.  In a sub-Penning mixture, we can rely on geometric properties of the scintillation cloud, which appear to offer much better position accuracy.  Here, we point out three alternative z-labeling methods for different drift geometries.   The two underlying geometries we will consider are parallel drift, where the anode and cathode are planar and the drift field is constant; and cylindrical drift, where the anode and cathode are concentric cylinders.   We will also discuss the multiple-volume case, where light can propagate past electrode structures from one drift field region to another.  

\paragraph {\textbf Parallel drift: truncated scintillation clouds} In a detector with parallel drift fields and a constant drift velocity $v_d$, we can gain some z-information by determining whether the (spherical, symmetric) scintillation cloud has been truncated by either the anode or the cathode.    If the scintillation electrons arrive symmetrically in time, including electrons extending to earlier and later times than the ionization cloud, then the event must have occurred in an open space in the middle of the detector.    An event occuring a distance $x$ from the anode will lack the tail of scintillation electrons from early times $t < x/v_d$ before the primary electrons.  An event occurring $x$ from the anode will be missing the tail of late electrons arriving $t > x/v_d$ after the primary electrons.    

This z-resolution is reliable in events where the ``missing'' tail is statistically significant.   An event with $N$ total expected scintillation electrons will be z-resolvable if it is roughly $z/\lambda < 0.5(log_{10}(N)-1)$ from either wall, where $\lambda$ is the photoconversion length.   If we choose $\lambda$ which is a large fraction of the z-dimension $z_{max}$ of the detector, then nearly all events are z-resolved.   Choosing $\lambda > z_{max}$  comes at a cost in scintillation photon statistics.   If we choose a photoconversion length much shorter than the detector size, i.e. $\lambda << z_{max}$, a region in the center of the detector will have unknown z-values {\em but} events near the anode and cathode will still be identifiable as such.   This corresponds well to one typical use of z-axis information, i.e. for performing fiducial volume cuts.  It is less useful for performing, e.g., z-dependent energy scale corrections.

\begin{figure}
\includegraphics[width=1.0\textwidth]{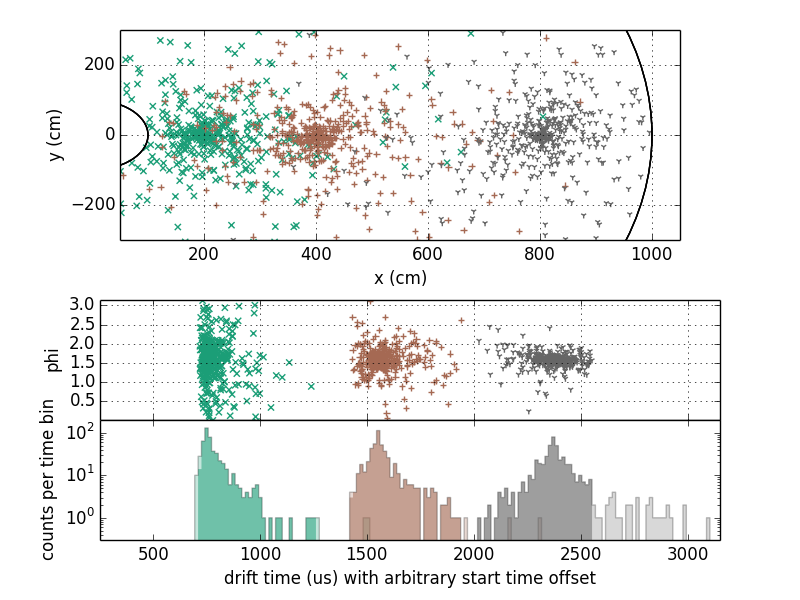}
\caption{Scintillation photoelectrons detected in $\phi$ and $t$ coordinates in a model cylindrical TPC with anode radius 100~cm, cathode radius 1000~cm, and a 200~cm scintillation attenuation length.  Top:  The initial-state photoelectron clouds are shown for three events with n$_{pe}$ = 500 and origins at $r_0=200$ (teal x), $r_0=400$ (brown +), and $r_0=800$ (grey y)~cm.   Primary particle ionization is not shown.   Botom: The electron detections are shown in ${\phi,t}$ projection, with arbitrary start-time offsets for each event for clarity.   The drift times are based on calculations for a high-pressure neon detector.   Note the (purely geometrical) effect of $r_0$ on the photoelectron $\phi$ distribution, and the arrival-time asymmetry which is partly geometric and partly due to the drift-velocity variation in $r$.  Fitting these distributions provides, independent of truncation, a good continuous measurement of $r_0$.  The histograms show scintillation photoelectron arrival times (dark colors) as well as extrapolations  (light colors) showing ``missing'' photoelectrons due to the cloud truncation by the anode and cathode.   The electron cloud from the $r_0=200$ event has its early tail truncated ($\Delta n_{pe}$=24) due to photons hitting the anode.  The electron cloud from the $r_0=800$ event is truncated ($\Delta n_{pe}$=38) at the cathode.   The $r_0=400$ event has statistically-significant 5-electron truncation at the cathode.  The truncations provide the measurement of $r_0$ which (as needed for fiducialization) is increasingly reliable closer to the anode and cathode surface.}\label{fig_r_phi}\end{figure}

\paragraph {\textbf Cylindrical drift: truncated and reshaped scintillation clouds} In a detector with cylindrical drift, as suggested for all of the high-pressure systems discussed, we have two geometric handles on the primary event's position $r_0$ along the drift coordinate $r$.   The cloud of photoconversion electrons is spherically symmetric, but we will detect it by a spatial projection onto cylindrical $z,\phi,t$ coordinates.   First, because the drift field varies with radius, $t$ is a nonlinear function of $r$; the spread and asymmetry of the electron arrival times can be used to determine $r$.  Second, the cloud's spatial extent $\lambda$ maps to a wider or narrower distribution of $\phi$ at larger or smaller $R_0$.   Examples of $\phi,t$ distributions for otherwise-identical events at different $r_0$ are shown in Fig. \ref{fig_r_phi}.   

\begin{figure}
\begin{center}
\includegraphics[width=0.8\textwidth]{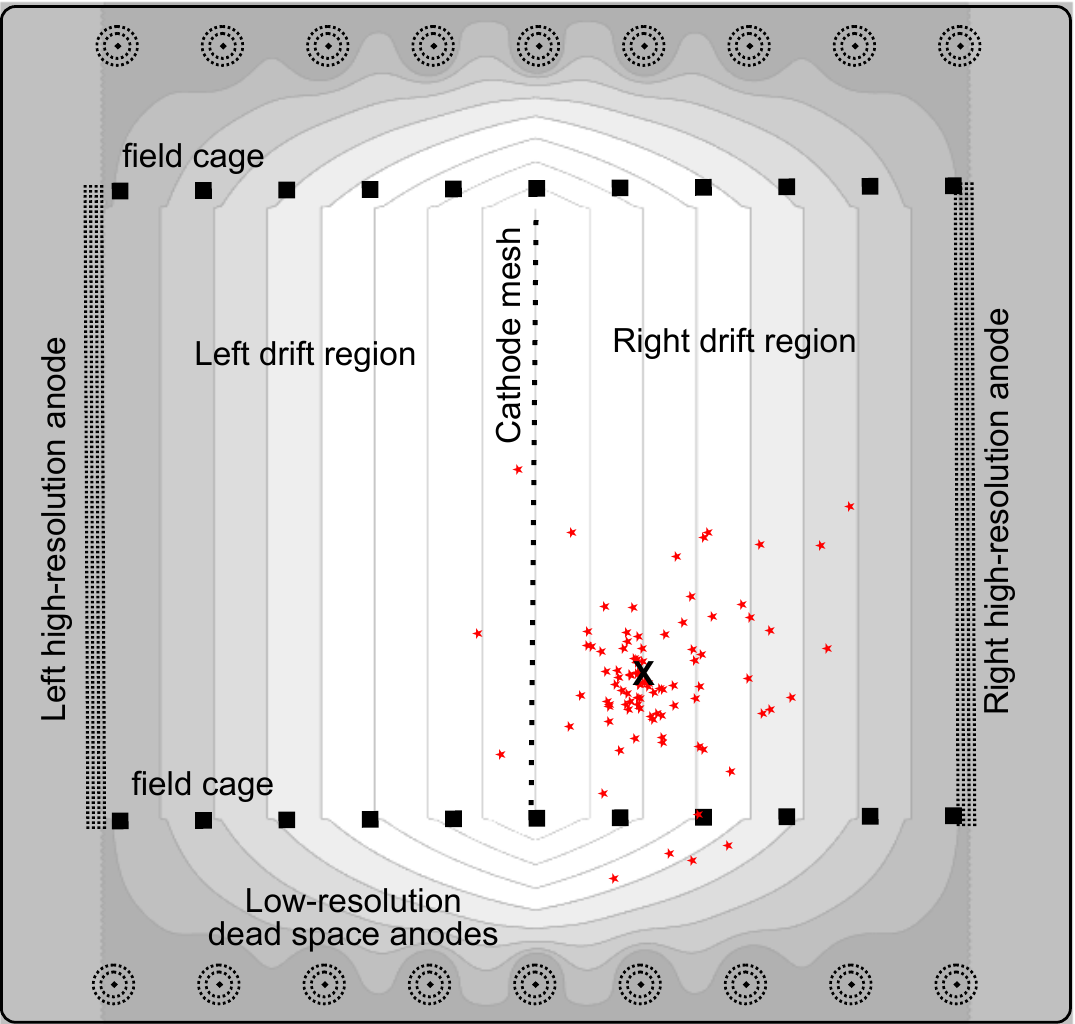}
\end{center}
\caption{Schematic of a two-sided TPC geometry showing how multiple drift regions contribute to event reconstruction in a sub-Penning gas.   Greyscale shows electric potential.   An event (black X) occuring in the right side of the TPC produces a spherically-symmetric cloud of photoelectrons (red stars).   The primary ionization drifts to the right anode along with most of the photoelectrons; if that was the only information, the event start time could not be determined since it is degenerate with the left/right drift-axis position.   In this picture, note that some scintillation photons have crossed the cathode, producing photoelectrons that drift left.   Comparison of the right-anode and left-anode electron distributions allows unambiguous knowledge of the ionization coordinate.   Also, four photons have converted in the ``dead region'' behind the field cage, which we show as instrumented with additional anodes.  Dead-region electrons are detected quickly at these anodes, and effectively provide an event start time.   Either the left/right or the dead region signals could suffice for reconstruction.}\label{fig_segmented}
\end{figure}

\paragraph {\textbf Multiple optically-contiguous drift volumes: positive event start time tags} In many conventional TPCs, there is a single drift volume (bounded by an anode, a cathode, and field-shaping electrodes); the region outside of the field-shaping electrodes is filled with the same working fluid but considered ``dead'' because its ionization does not drift towards the main anode plane.   With a sub-Penning TPC, some photoelectrons will convert in these dead volumes; by instrumenting some or all dead volume as a low-granularity, short-drift TPCs, we can essentially obtain a fast ``prompt light'' tag on all events sufficiently close to any wall.   The geometry is illustrated in Fig. \ref{fig_segmented}.  (Note that the dead-space instrumentation has the same optical-baffling requirement as the main anode plane.) Similarly, many TPCs use bidirectional drift: a mesh cathode plane runs through the center of the detector, and two opposite walls are instrumented as anode planes.   In a bidirectional sub-Penning TPC, an event occurring in the left half-volume can send scintillation photons through the cathode and into the right half-volume.  The direct ionization and most of the photoelectrons are detected on the left anode plane, and a handful of photoelectrons are detected on the right.   Combining the left and right arrival time distributions allows us to reconstruct the event start position and time.

\subsection{Scintillation timing pulse shape discrimination}
In many applications, we need to discriminate high-dE/dx events (nuclear recoils, alpha decays) vs. low-dE/dx events (beta/gamma events)  even when the relevant tracks are spatially unresolved.   A sub-Penning detector has access to the ionization/scintillation ratio, since both components are detected efficiently.   It is also known, however, that different scintillation decay times are generated at different ionization densities, and in many cases scintillation timing PSD is very powerful at signal/background separation.   Nuclear recoil identification has been studied in liquid neon \cite{Lippincott:PhysicalReviewC:2012} and argon \cite{Lippincott:PhysRevC:2008} and $\alpha$-$\beta$ discrimination is used in liquid scintillators \cite{Back:NuclearInstrumentsAndMethodsInPhysicsResearch:2008}.    

In a sub-Penning TPC, we can obtain some scintillation timing information because the primary ionization has a ``head start'' over the slow scintillation photo lectrons.   The applied $\vec{E}_{drift}$ begins acting on the ionization electrons immediately, but the excited atoms remain at rest during the scintillation decay time, and the scintillation photoelectrons will arrive late.   A long scintillation decay time will cause a spatial offset between the photoelectron cloud (whose centroid can be found with precision $\sim \lambda/\sqrt{N_{pe}}$) and the primary ionization.   In ``fast'' scintillators (Xe, CF$_4$, organic liquids) this time offset is negligible compared to any reasonable drift velocity.  However, in slower scintillators Ar ($\tau_{slow}=1.5 ~\mu$s), Ne ($\tau_{slow}=15 ~\mu$s), and He (very slow), the offset may be detectable as long as $\lambda$ is small.

\begin{figure}
\includegraphics[width=1.0\textwidth]{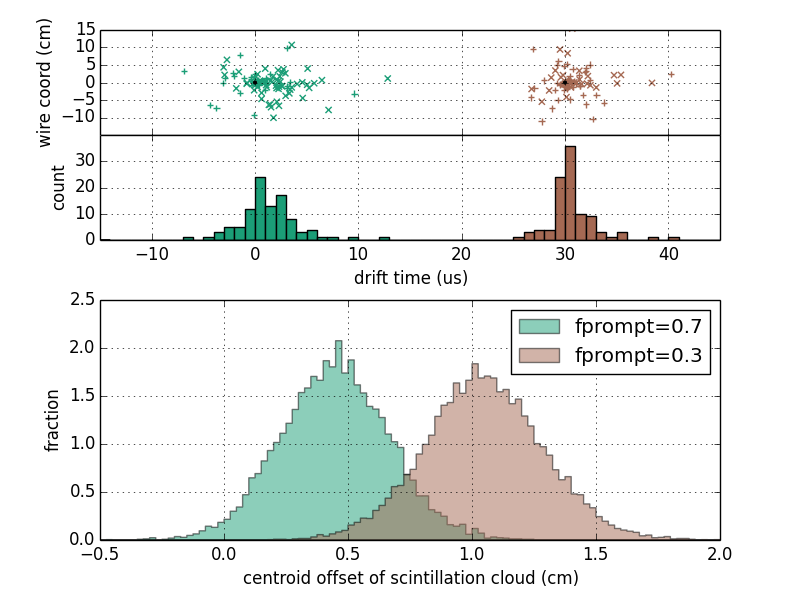}
\caption{Scintillation pulse shape discrimination capabilities of a sub-Penning TPC with linear drift.  In each case the teal (left) event is a nuclear recoil, the brown (right) event is an electron.  Top: two example events at 100~pe each in liquid Ar:TEA under high (5~kV/cm) drift fields.    The earlier (teal) event has $F_{prompt}=0.3$, characteristic of a nuclear recoil.  The later (brown) event has $F_{prompt}=0.7$, characteristic of an electronic event.   The time delay between the scintillation cloud and the ionization (black dots) is clearly visible both in the projections and in the time histogram.  Bottom: Ionization/scintillation centroid separation as a recoil/electron pulse shape discriminant.   $F_{prompt}=0.3$ and  $F_{prompt}=0.7$ have cleanly separable distributions for $n_{pe}=100$ events.  Additional discrimination power would be obtained from the ionization/scintillation ratio.}\label{psd}

\end{figure}

In liquid argon at dark-matter-like energies, nuclear recoils typically lead to $\sim$70\% slow (1.5~$\mu$s) scintillation, while $\beta/\gamma$ events produce $\sim$30\%  \cite{Lippincott:PhysRevC:2008}.  Drift velocities of $\sim$3~mm/$\mu$s are obtainable in high drift fields (5~kV/cm) \cite{Walkowiak:NuclearInstrumentsAndMethodsInPhysicsResearch:2000}; while high fields quench the scintillation amplitude,  there is no evidence that they alter the PSD properties \cite{Cao:PhysicalReviewD:2015}.    In order to measure an offset between the scintillation photoelectron cloud and the ionization cloud, we need the scintillation photoconversion length to be short.    In Fig. \ref{psd} we show the PSD capabilities of a detector with an ``unusually high'' TEA concentration and 5~cm photoconversion length.   (Note that this limits the range of the scintillation-based z-axis reconstruction described above.)

\section{Four large detector case studies: noble gases} 

Over the past 20 year physicists have developed successful, stable detector technologies suitable for making keV-threshhold, tonne-scale detectors in argon, neon, and xenon; 100-keV, multi-kT-scale detectors in argon and hydrocarbons, and 10-MeV, MT-scale detectors in water.   Is it worthwhile to push the development of a \emph{particularly challenging}, narrowly-stable, unproven gas detector technology, in order to work with substantially the same list of target nuclei?   We believe the answer is ``yes'', in particular, in order to create opportunities for science instruments at far larger scales than those considered today.   In this section, we present four sketches of ultra-large detectors which we believe illustrate some of the possible directions of opportunity.   

In each of the cases below, we (a) suggest an appropriate scale and target mass for this choice of gas and deployment, (b) comment briefly on data/signal/background issues peculiar to the use of the sub-Penning medium, and (c) argue that these issues could be compatible with (or could improve on) the current state-of-the-art in event reconstruction with similar physics goals.   Most conventional detector-design considerations (materials, radioactivity budgets, readout details) are neglected unless we feel there is a sub-Penning-specific aspect worth commenting on.   For the specific engineering complexities of solution-mined salt caverns, we refer the reader to  \cite{Monreal:EprintArxiv14100076:2014}.  While acknowleding that the concept is far from proven, we hope that these detector sketches provide motivation for further engineering studies on those caverns as well as on the gases themselves.

\begin{table}
\begin{center}
\begin{tabular}{|c|c|c|c|c|c|}
\hline
Case & Gas & Physical properties & Drift geometry, time& Mass & Application \\
\hline
1 & Xe:TMAE & 3~m$\oslash\times$1.5~m liquid & vertical, 100~kV, 700~$\mu$s & 30~t & WIMPs \\
2a & Xe:TMAE & 3.6~m$\oslash\times$7~m, 55~bar & bidirectional, 150~kV, 7~ms& 32~t & 0$\nu\beta\beta$\\
2b & Xe:TMAE & 16~m$\oslash\times$30~m, 1~bar & multi-anode & 32~t & 0$\nu\beta\beta$\\
3 & Ne:Xe & 16~m$\oslash\times$65~m,100~bar & 0.5~m$\oslash$ anode, 200~kV, 8~ms & 1~kt & WIMP, solar $\nu$\\
4 & (Ar:Xe):TMA & 55~m$\oslash\times$15~m, liquid & vertical, 1.5 MV, 20~ms & 50~kT & Long baseline \\
5 & H$_2$:TEA & 50~m$\oslash\times$250~m,200~bar & multi-anode & 18~kT & reactor/geo/SN$\nu$\\
\hline
\end{tabular}
\end{center}
\caption{Summary table: examples of giant TPCs possible with sub-Penning mixtures.  Motivation for these dimensions and pressures is given in the text.  We suggest some drift configurations based on Magboltz \cite{Biagi:1999nwa} calculations.  Note the long drift times, which imply high stringent purification requirements.  Cases 2b (slow gas) and 5 (very long distances) would have unreasonable drift times if instrumented as a single volume and might work better if subdivided.}\label{summary_table}
\end{table}

\begin{figure}
\begin{center}
\includegraphics[width=0.7\textwidth]{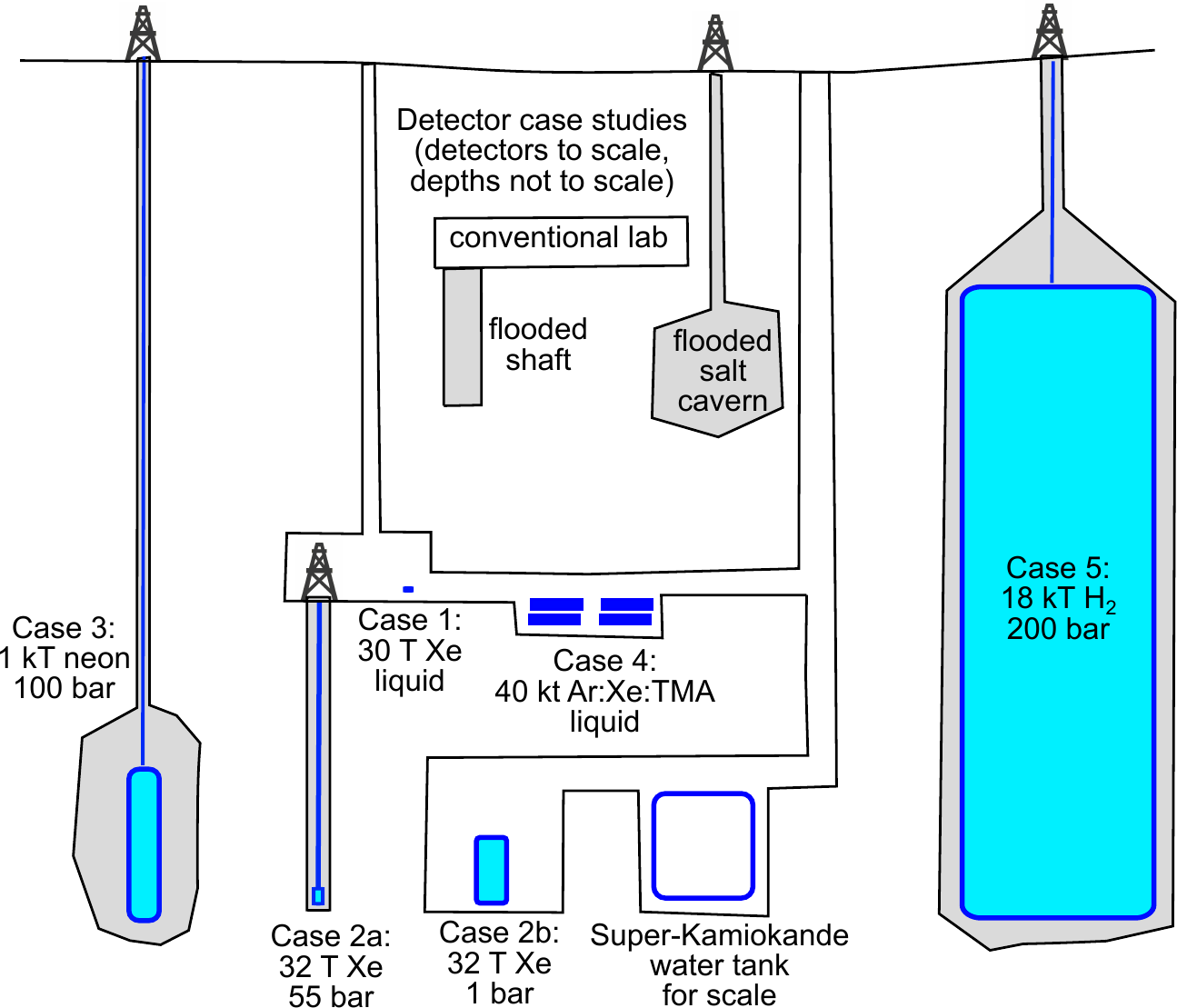}
\end{center}
\caption{Summary sketch of detector case studies from Table \ref{summary_table}.   Detectors are shown to scale, but depths are arbitrary.  Super-Kamiokande is shown for scale.}\label{overview_drawing}
\end{figure}

\subsection{Case 1: a 30~tonne Xe:TMAE two-phase TPC for dark matter} 

The high density of liquid xenon allows the construction of compact, thoroughly self-shielded detectors appropriate for dark matter searches.    We note that dark matter detectors like PANDAX, XENON1T and LZ rely critically on photomultiplier tubes.  PMTs are expensive, high in radioactivity, and even at their best are of fairly coarse granularity and low quantum efficiency.    By comparison, TMAE appears to offer a 50\% photoelectron yield.   Compare this to a  5--10\% net scintillation detection efficiency (including the effect of reflections, photocathode coverage, and QE) in the LZ conceptual design \cite{Collaboration:EprintArxiv150902910:2015}.  

By boosting the statistics of scintillaton light detection, we boost the $\gamma$/nuclear recoil separation power, and extend its use to much lower threshholds.   LZ's S1/S2 discrimination begins failing below 4~keVnr when the mean number of scintillation photoelectrons falls to ~0--1.   A TMAE photoconversion system at 50\% QE would retain few-photoelectron detection capabilities down to $\sim1$~keVnr with a 2~pe threshhold. Note that fiducial cuts become inefficient at 2~pe, since an analysis requires the presence of both ``leading'' and ``trailing'' photoelectrons in order to establish a minimum separation from the walls.   Scintillation-geometry-based fiducialization is impossible with 1~pe, but fiducialization based on diffusion of the primary signal may be good enough.    

\subsection{Case 2: a 32~tonne Xe:TMAE gas TPC for neutrinoless double beta decay and dark matter} 

There are many advantages to doing rare-event search experiments in {\em liquid} noble gases rather than gases: a dense liquid has good stopping power for self-shielding against gamma ray backgrounds, and high density minimizes the physical size and mass of readout planes, shielding tanks, etc..  For the special case of neutrinoless double beta decay, we find several advantages to returning to the gas phase, as shown by the NEXT collaboration \cite{Collaboration:Jinst7:2012}.   First, in the gas phase, the Fano factor is very low (f$\sim$0.15), allowing us to obtain very high energy resolutions from ionization electron counting, whereas in liquids it is high (f$\sim$20, although some of this may be recovered from ionization/scintillation correlations.)   The low Fano factor is available up to a density of $\sim$0.5~g/cm$^3$ \cite{bolotnikov1999studies}, above which there is a gradual transition to liquid-like resolutions.   Second, both double-beta-decay events and beta/gamma backgrounds are frequently ``pointlike'' and unresolved in a liquid.  In a lower density gas, both event types may be resolvable as extended ionization tracks whose topology is a useful background-rejection handle.   However, the technique has only been used with detectors of fairly small target mass.   We argue that a sub-Penning xenon mixture could enable the construction of a very large high pressure Xe $0\nu\beta\beta$ experiment.

\subsubsection{Case 2a: high pressure} Consider a high-pressure xenon vessel or balloon 3.6~m in diameter and 7~m high with a gas density of 500 kg/m$^3$, i.e. at about 55~atm, yielding a 32~t xenon target with reasonably good self-shielding.   This high pressure could be obtained without a large pressure vessel by: 
\begin{enumerate}\item Placing a light pressure vessel at the bottom of a conventionally-mined vertical shaft and flooding it with water.  The existing \#4 and \#6 winzes at Sanford Lab are, in the absence of a dry 7400L laboratory, long enough.  (At 3.7~m and 4.4~m wide, respectively, they are in principle large enough for a 3.6~m diameter vessel, but additional space for shielding would be wanted in practice.)

\item Placing an inflatable xenon balloon in a solution-mined salt cavern as discussed in  \cite{Monreal:EprintArxiv14100076:2014}.   An example of the successful construction and filling of a thin-balloon detector can be found in  \cite{Benziger:NuclearInstrumentsAndMethodsInPhysicsResearch:2007}.  In this case, it is easy to obtain many meters of open space between the xenon and the cavern walls, which for shielding purposes could be filled with propane.  Propane is a good density match to xenon at 55~bar and relieves pressure gradients from the xenon balloon.
\end{enumerate} 

To operate this as a cylindrical-drift TPC, consider the vessel as a grounded cathode.  A central anode cylinder with a 25~cm radius, hanging freely in the center of the gas vessel, carries a triple-GEM system backed by a SIPM photoluminescent readout array of 10~m$^2$ area.  The radial E-field variation introduces spatial nonunformities in the energy calibrations.  To operate with linear-drift, we might opt for bidirectional drift in two 3.5~m-high drift volumes separated by a cathode  mesh held at negative high voltage.   This requires somewhat more anode plane area (20~m$^2$) and field-shaping hardware and thus higher radioactivity.   Either system in principle allows NEXT-like energy resolution, although it is not clear whether NEXT-like track spatial reconstruction is possible at these sizes.

\subsubsection{Case 2b: low pressure} As an unusual alternative, note that 32~t Xe at \emph{atmospheric pressure} occupies only 6000~m$^3$, which would fit comfortably in the large existing caverns at SNOLAB, Kamioka, Soudan, or Gran Sasso.  This removes the unconventional engineering challenges associated with large pressure vessels or hydrostatically-pressurized deployments, at the cost of (a) no self-shielding and (b) a very large instrumented volume.   However, the sub-Penning approach lends itself well to instrumenting such a large volume: no area-scaling PMT array is needed, no cryostat is needed, and the DAQ channel density can scale with the gas density.   

\subsection{Case 3: a 1~kT Ne:Xe or Ne:H$_2$ detector for dark matter and solar neutrinos} 

The use of neon as a detector material is extremely attractive from the point of view of radiopurity and cost.   Neon has no long-lived radioisotopes, and with a boiling point far below the freezing point of typical contaminants, it can be ultrapurified very thoroughly.   Neon thus affords something very close to a zero-background medium for low-energy solar neutrino physics.   A neon proportional counter's high energy resolution (better than water \cerenkov) and directionality (better than liquid scintillator) would make it an excellent target for medium-energy neutrino oscillation studies \cite{Bungau:PhysicalReviewLetters:2012,adelmann2014cyclotrons} or neutrino-based indirect dark matter searches \cite{Rott:EprintArxiv151000170:2015}.   Proposals for neon-based experiments have included CLEAN (135~t liquid neon, scintillation readout) \cite{McKinsey:AstroparticlePhysics:2005},  SIGN (1~t gaseous neon at 100~bar, WLS fiber readout) \cite{white2007sign}, and a neon mode for MiniCLEAN (500 kg liquid neon, scintillation readout) \cite{Rielage:PhysicsProcedia:2015}.

Neon's utility for spin-independent WIMP searches is weakened, but not destroyed, by its low atomic number.  In the zero-threshhold limit, compared to Xe, Ne has 4\% the event rate per kg, and even worse even rate per m$^3$ (although this varies with temperature and pressure).   However, the event rates \emph{per unit target cost} are very nearly comparable.  Neon is particularly suitable for low-mass WIMP searches, and large-but-plausible target masses and volumes could give neon competitive sensitivity at all masses.  We wish to explore the possibility that (a) neon's low cost makes large \emph{masses} practicable, and (b) the sub-Penning gas mixture approach makes large \emph{volumes} practicable.

\begin{figure}
\begin{center}
\includegraphics[width=0.6\textwidth]{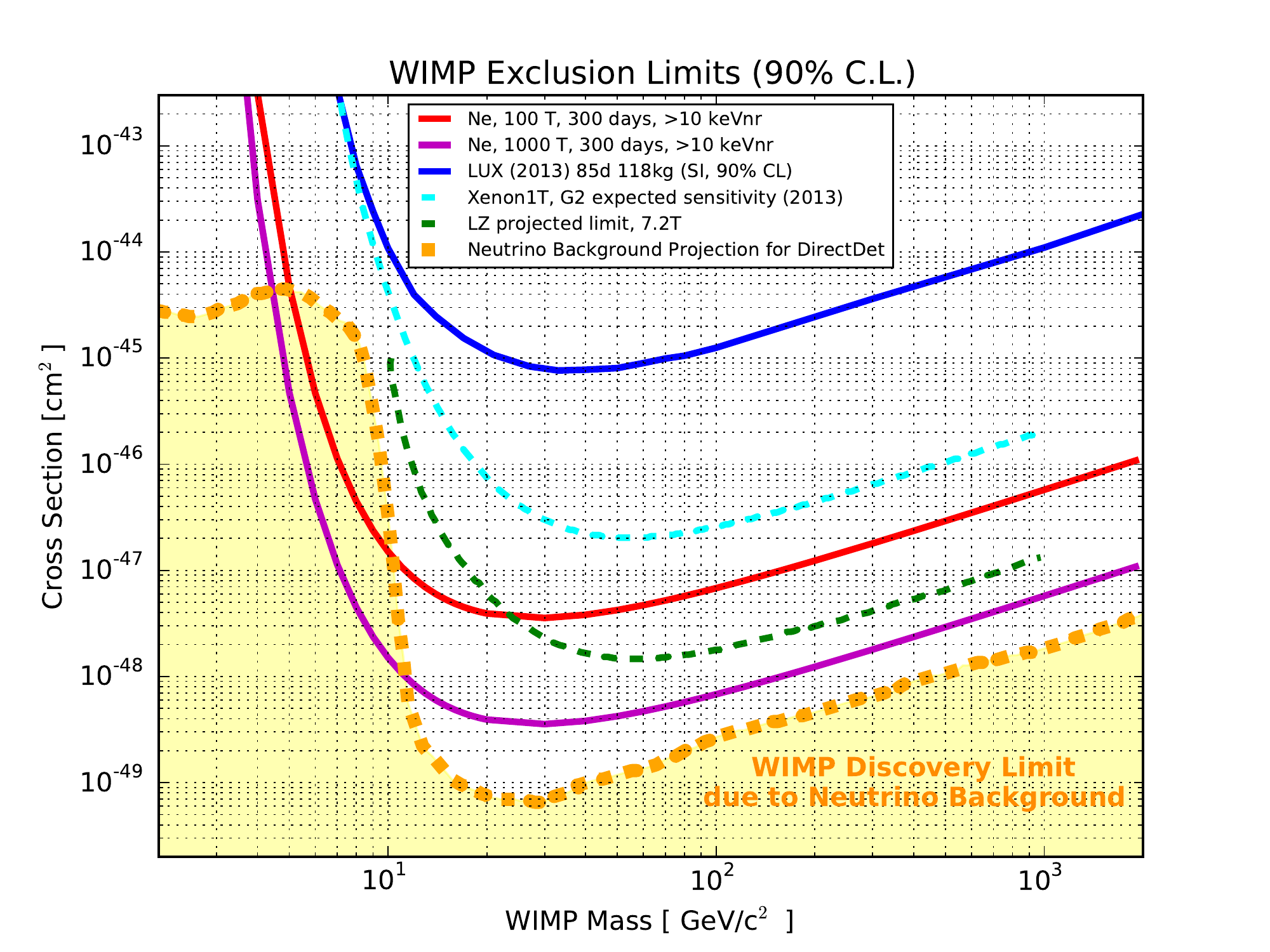}
\end{center}
\caption{WIMP discovery potential for 100~t and 1~kt Ne-based dark matter experiments, calculated for a 10~keVnr threshhold, 300 d livetime, and 2-event requirement for discovery.  Curves are from DMTOOLS \cite{dmtools}.} \label{fig_ne}
\end{figure}

A two-phase target would require scaling-up of conventional technologies, along the lines of case 1 but much larger and with lower-temperature cryogenics requirements.  However, a solution-mined salt cavern, or a hydrostatically-pressurized conventional excavation, may offer a plausible route to an ultralarge single-phase detector.   Consider a neon balloon of diameter 16~m and height 80~m, at a depth of 1000~m and pressurized to 100~bar (all of which are routine parameters for gas storage in salt caverns).   This is straightforward to instrument as a cylindrical drift chamber by hanging a 50~cm diameter anode cylinder in the center, and grounding the balloon.  All of the high voltage and DAQ instrumentation is confined to the anode, which can be designed to fit down the cavern-access borehole in one piece.    Containing about 1000~t of gaseous neon, the dark-matter sensitivity of such a detector in a 300-day run would reach coherent-neutrino-background sensitivity limits for $5 < M_\chi < 11$~GeV (see Fig. \ref{fig_ne}) and outperform LZ by a factor of 3 at larger $M_\chi$.   A 1000~t target would see 2700 solar neutrinos, including 1800 $pp$ neutrinos, per day.

\subsection{Case 4: Ar:Xe:TMA as a long-baseline neutrino target} 

The DUNE and LNBO programs involve the construction of large liquid-argon TPCs.  The GLACIER detector concept, in particular, depicts a two-phase detector with very long drift lengths; we use this as the basis for discussing the uses of an argon-based sub-Penning mixture.   One approach to using sub-Penning mixtures in large argon detector would be to choose an Ar:TMAE or Ar:TEA Penning mixture, which would provide a large photoionization cloud based on the argon scintillation at 127~nm.   More powerfully, consider the following \cerenkov-sensitive three-component mixture:  In Ar:Xe 99.9:0.1, the 127~nm argon eximer scintillation is almost entirely quenched, and VUV scintillation light appears only at the 174~nm wavelength associated with xenon.   Meanwhile, this mixture has good transparency from down to 145~nm and a band of \emph{reasonable} transparency around 133~nm \cite{Neumeier:Epl:2015_2}.    Therefore, an Ar:Xe mixture can propagate VUV \cerenkov radiation in the bands 130-136~nm and 145--160~nm which are uncontaminated by scintillation light.   By doping with TMA  (threshhold ~147~nm), we can photoconvert \cerenkov photons from the bluer band and detect the photoelectrons.   Despite the narrow transparent band, the \cerenkov yield is acceptable (up to $\sim$90~pe/cm).   To increase the yield, one would search for a dopant that is ionized by the 145--160~nm photons without responding to the tail of the scintillation spectrum.  (Judging by its low-pressure properties, TEA may be appropriate, but its properties in solution in LAr are not well known.)

For high-energy events, cone reconstruction would provide useful particle-ID information in hadronic events.  For example, in searches for the proton decay $p^+\rightarrow K^+ \nu$,  a neutrino-induced single $\pi^+$ can sometimes fake the $K^+$ in liquid argon.  In an (Ar:Xe):TMA detector, the $\pi^+$ is above \cerenkov threshhold while the $K^+$ is below.   At lower energies, cone reconstruction could measure the directions of events, like supernova neutrinos, whose electron tracks are unresolved (or have a head-tail ambiguity) in ionization data alone.

\section{Hydrogen-rich sub-Penning mixtures for antineutrino physics? H$_2$:TEA and  CH$_4$:TMAE}

The tagged antineutrino capture reaction $\bar{\nu_e} + p \rightarrow n + e^+$, the basis of almost all reactor neutrino and geoneutrino detection, requires a hydrogen-rich detector material.   To date, that has restricted the field\footnote{The MUNU and Savannah River experiments are exceptions.} to using liquid scintillator arrangements for detecting the $e^+$, with experiments varying only in different approaches to segmentation and neutron capture.   Thus, reactor neutrino physics has been tied closely to (a) scintillator-like energy resolution and (b) cost and size scales driven by PMTs.   If sub-Penning gas TPCs offer a route towards ultra-large next generation detectors at all, is there an appropriate hydrogen-rich gas?    The two likely candidates are H$_2$ and CH$_4$.   In both cases, we run into apparent gaps in the spectroscopy literature that prevent confident pursuit of the point:   

\begin{enumerate}
\item \textbf{Do H$_2$ or CH$_4$-rich mixtures scintillate?}   H$_2$ is a well-attested but fairly rarely used drift chamber gas.   It is inherently free of many cosmogenic backgrounds ($^8$He, $^9$Li, $^{11}$C) and easily stripped of U/Th/Kr contaminants.  The literature is nearly silent on dense or liquid H$_2$'s VUV scintillation properties.  However, one source  \cite{Dieterle:NuclearInstrumentsAndMethods:1979} reports a dim scintillation yield of 15~photons/MeV between 185--210~nm.   No experiments were done at the more-difficult shorter wavelengths and no detailed spectrum is available.   Solid H$_2$ does not scintillate \cite{Stenum:TheJournalOfChemicalPhysics:1993}.    If the 15~photons/MeV number is correct, then H$_2$:TMAE is probably an ``adequate'' sub-Penning gas at the energy scales needed for inverse beta decay detection.  If additional scintillation is present below 185~nm then other mixtures would also work.   An H$_2$ target would be entirely free from $^{12}$C spallation products ($^9$Li, $^8$He, $^{11}$C, etc.) which are some of the most difficult backgrounds in scintillator detectors.

Pure CH$_4$ is known not to scintillate \cite{Pansky:NuclearInstrumentsAndMethodsInPhysicsResearch:1995}, but it is possible that CH$_4$ could be doped with a scintillating gas.  A likely candidate is CF$_4$, a bright scintillator with both optical components and 160~nm UV components.  If the UV scintillation survives in a CF$_4$:CH$_4$ mixture, then (CH$_4$:CF$_4$):TMAE could be used as a hydrogen-rich sub-Penning gas.  VUV Ar scintillation is quenched by CH$_4$ \cite{Jones:2013mfa}. 


\item \textbf{Is liquid CH$_4$ transparent to TMAE-ionizing radiation?}  CH$_4$ has been demonstrated successfully as a liquid-phase TPC medium \cite{Aprile:1987jt,Tanaka:2004br}.  It is the lowest-cost commodity hydrocarbon by a wide margin, often trading at around \$100/t.  In the absence of scintillation, pure liquid methane is a \cerenkov medium, VUV \cerenkov photons could serve as the TMAE-ionizing radiation.  Is this practical?  Again, experimentally-unanswered spectroscopy questions get in the way: first, what are the TMAE threshold and yield when dissolved in CH$_4$?  Is it 210~nm, as in the gas phase, or closer to ~270~nm as seen in solution in liquid argon \cite{hitachi1997photoionization}?   Second, is liquid methane transparent at the relevant threshold?   Gaseous methane is transparent to wavelengths longer than 145.5~nm\footnote{Gas-phase cross sections \cite{mainz_spectral_atlas} imply, at liquid density, a 10~cm UV absorption length at 152~nm} but no actual liquid data appears in the literature between 134~nm and 400~nm.   

Although some heavier hydrocarbons in the gas phase (alkanes, ethers) are known or likely transparent in the 210--270~nm range, it is not clear whether this carries over to the liquid state.  It would be particularly interesting if a \cerenkov-sensitive sub-Penning mixture like C$_3$H$_8$:TMAE were possible in a room-temperature liquid hydrocarbon TPC  \cite{Holroyd:NuclearInstrumentsAndMethodsInPhysicsResearch:1985}.   
\end{enumerate}

\subsection{Why use TPCs for antineutrino physics?} \label{sec_why_h2_tpcs}
The inverse beta decay signal is slightly different in a drift chamber than in a typical large scintillator detector \cite{dawson_2014}.    The reaction $\bar{\nu_e} + p \rightarrow n + e^+$ actually results in four separate energy deposits in a detector: first, the $e^+$ kinetic energy (KE$_e$), which is equal to the neutrino kinetic energy E$_\nu$ minus 1.8~MeV and appears as a single track.  Second and third, two 511~keV $\gamma$s from the positron annihilation, almost always seen as a chain of Compton scatters.  Fourth, a delayed energy deposit (for $n+H \rightarrow d +\gamma$, also a chain of Compton scatters) from capture of the neutron.   There will also be small nuclear-recoil energy deposits from the neutron thermalization.  As a high-resolution detector capable of separating all of these components, a TPC stands in contrast to unsegmented scintillator detectors\footnote{A detector with extreme segmentation, particularly a Raghavan Optical Lattice, can separate the primary positron from the annihilation gammas \cite{Rountree:ScienceAndTechnologyOfALowEnergySolar:2010} like this.} where the first three deposits are detected as a single ``prompt'' flash with energy $E_p = E_\nu - 0.7$~MeV.   In such detectors the delayed neutron-capture energy is the main background-rejection handle, while in a TPC, the main background-rejection handle is the four-hit topology.  (The neutron capture delay time might also be visible, at least statistically, if the electron drift velocity is sufficiently fast compared to the neutron diffusion and capture parameters.) 

We discuss the four-hit topology in order to note the key benefit of TPC antineutrino detection.   In a scintillator detector, the ``prompt'' signal is the sum of KE$_e$ and the 1022~keV positron annihilation.  This means that the energy resolution has the form $\Delta  \mathrm{KE}_e = a \sqrt{1.022 \mathrm{MeV} + \mathrm{KE}_e}$, with $a$ being the intrinsic detector energy uncertainty, typically with $a > 0.03 \sqrt{\mathrm{MeV}}$.  If the KE$_e$ deposit is measured and spatially separated from the 511~keV annihilation blobs, then the energy uncertainty is $\Delta  KE_e = a \sqrt{\mathrm{KE}_e}$, a dramatic improvement in energy resolution for low-energy neutrinos.   This could be important in particular for reactor neutrino mass hierarchy determination \cite{PhysRevD.87.033005}, where energy resolution near threshold is crucial.    

Beyond the question of finding a suitable gas mixture, many detector-size and engineering questions appear daunting.  In case of H$_2$, this is due to the low gas densities obtainable.  Consider Case 5, an 18~kt H$_2$ target (proton count equivalent to 125~kt scintillator, and capable of seeing $\sim 6$ inverse beta decay events from a supernova in Andromeda \cite{doi:10.1146/annurev-nucl-102711-095006}) could be obtained by filling an extremely large salt cavern (80~m$\oslash\times$240~m) with H$_2$ at 200~bar.   This is geotechnically plausible, since salt caverns of this span and pressure already exist, but clearly unusual in scale as seen in Fig. \ref{overview_drawing} and Table \ref{summary_table}.  CH$_4$ in the same cavern would have up to twice the proton count, depending on the CF$_4$ mixing ratio.  For a seafloor H$_2$-based geoneutrino detector with Hanohano-like (10~kt scintillator) cababilities \cite{Learned:2007zz}, we might want a 1.5~kt H$_2$ detector at 5~km depth (500~bar pressure), but this results in a vessel the size of SuperKamiokande.   Liquid methane is very nearly a one-for-one replacement of liquid argon in two-phase\footnote{Single-phase detectors are not usually sensitive to isolated electrons and would not see the photoelectron cloud at all.} detectors like GLACIER \cite{Rubbia:09081286:2009}, but the mine-safety engineering challenges might be prohibitive.   

\section{Conclusions}

We show that a new class of gas mixtures, corresponding roughly to ``underquenched'' Penning mixtures, could be used as the working medium in new ultra-large time projection chambers.   In a large open TPC volume, directly-ionized electrons and scintillation/photoionization electrons would be independently detectable due to their different spatial distributions.  If some photons can reach the detector walls before photoionizing, it confers the ability to reconstruct the full 3D position of an event.   Optically-baffled gain structures should permit stable gas operation, despite these mixtures' potential for avalanche feedback in conventional counters.   The new mixtures appear to provide realistic options for gas TPCs of unprecedentedly large size, with options for useful detectors based on xenon, argon, neon, hydrogen, and possibly methane.

\section{Acknowledgements}

The authors thank Janet Conrad, Vic Gehman, Dan McKinsey, Harry Nelson, Mike Witherell, and Jaret Heise for information and useful discussions.  This work was supported by the Department of Energy under DE-SC0013892.

\bibliographystyle{hunsrt_bm}
\bibliography{coptpc}

\begin{thebibliography}{10}

\bibitem{Vavra:NuclearInstrumentsAndMethodsInPhysicsResearch:1996}
J.~Va'vra, J.~Kadyk, J.~Wise, and P.~Coyle.
\newblock Study of photosensitive mixtures of {TMAE} and helium, hydrocarbon or
  {CF$_4$}-based carrier gases.
\newblock {\em Nuclear Instruments and Methods in Physics Research Section A:
  Accelerators, Spectrometers, Detectors and Associated Equipment},
  370(2-3):352--366, 2 1996.

\bibitem{Alvarez:JournalOfInstrumentation:2014}
V.~\'{A}lvarez {\em et~al.}
\newblock Characterization of a medium size {Xe/TMA} {TPC} instrumented with
  microbulk {M}icromegas, using low-energy $\gamma$-rays.
\newblock {\em Journal of Instrumentation}, 9(04):C04015, 4 2014.

\bibitem{Holroyd:NuclearInstrumentsAndMethodsA:1987}
R.~A. Holroyd, J.~M. Preses, C.~L. Woody, and R.~A. Johnson.
\newblock Measurement of the absorption length and absolute quantum efficiency
  of {TMAE} and {TEA} from threshold to 120 nm.
\newblock {\em Nuclear Instruments and Methods A}, 261:440--444, 1987.

\bibitem{Nakagawa:NuclearInstrumentsAndMethodsInPhysicsResearch:1993}
K.~Nakagawa, K.~Kimura, and A.~Ejiri.
\newblock Photoionization quantum yield of
  {TMAE}(tetrakis-dimethylaminoethylene) in supercritical xenon fluid.
\newblock {\em Nuclear Instruments and Methods in Physics Research Section A:
  Accelerators, Spectrometers, Detectors and Associated Equipment},
  327(1):60--62, 3 1993.

\bibitem{hitachi1997photoionization}
A.~Hitachi {\em et~al.}
\newblock Photoionization quantum yields of organic molecules in liquid argon
  and xenon.
\newblock {\em Physical Review B}, 55(9):5742, 1997.

\bibitem{Dieterle:NuclearInstrumentsAndMethods:1979}
B.~Dieterle, P.~Denes, and J.~B. Donahue.
\newblock A scintillating gaseous hydrogen drift chamber for use as a
  position-sensitive target.
\newblock {\em Nuclear Instruments and Methods}, 165(2):351--353, 10 1979.

\bibitem{Neumeier:Epl:2015}
A.~Neumeier {\em et~al.}
\newblock Intense vacuum ultraviolet and infrared scintillation of liquid
  {Ar-Xe} mixtures.
\newblock {\em EPL (Europhysics Letters)}, 109(1):12001, 1 2015.

\bibitem{Rubin:2013Jinst8P08001:2013}
A.~Rubin {\em et~al.}
\newblock Optical readout: a tool for studying gas-avalanche processes.
\newblock {\em 2013 JINST 8 P08001}, 5 2013.

\bibitem{Dafni:NuclearInstrumentsAndMethodsInPhysicsResearch:2009}
T.~Dafni {\em et~al.}
\newblock Energy resolution of alpha particles in a microbulk micromegas
  detector at high pressure argon and xenon mixtures.
\newblock {\em Nuclear Instruments and Methods in Physics Research Section A:
  Accelerators, Spectrometers, Detectors and Associated Equipment},
  608(2):259--266, 9 2009.

\bibitem{bondar2002high}
A.~Bondar, A.~Buzulutskov, and L.~Shekhtman.
\newblock High pressure operation of the triple-{GEM} detector in pure {Ne},
  {Ar} and {Xe}.
\newblock {\em Nuclear Instruments and Methods in Physics Research Section A:
  Accelerators, Spectrometers, Detectors and Associated Equipment},
  481(1):200--203, 2002.

\bibitem{Sood:NuclearInstrumentsAndMethodsA:1994}
R.~K. Sood, Z.~Ye, and R.~K. Manchanda.
\newblock Ultra-high pressure proportional counters part ii. xenon.
\newblock {\em Nuclear Instruments and Methods A}, 344:384--393, 1994.

\bibitem{Aston:NuclearInstrumentsAndMethodsInPhysicsResearch:1989}
D.~Aston {\em et~al.}
\newblock Development and construction of the {SLD} {C}herenkov ring-imaging
  detector.
\newblock {\em Nuclear Instruments and Methods in Physics Research Section A:
  Accelerators, Spectrometers, Detectors and Associated Equipment},
  283(3):582--589, 11 1989.

\bibitem{Collaboration:Jinst7:2012}
{NEXT Collaboration} {\em et~al.}
\newblock {NEXT-100} technical design report ({TDR}). executive summary.
\newblock {\em JINST 7}, 2 2012.

\bibitem{Lewis:NuclearInstrumentsAndMethodsInPhysicsResearch:2015}
P.~M. Lewis {\em et~al.}
\newblock Absolute position measurement in a gas time projection chamber via
  transverse diffusion of drift charge.
\newblock {\em Nuclear Instruments and Methods in Physics Research Section A:
  Accelerators, Spectrometers, Detectors and Associated Equipment}, 789:81--85,
  7 2015.

\bibitem{Lippincott:PhysicalReviewC:2012}
W.~H. Lippincott {\em et~al.}
\newblock Scintillation yield and time dependence from electronic and nuclear
  recoils in liquid neon.
\newblock {\em Physical Review C}, 86(1):15807, 7 2012.

\bibitem{Lippincott:PhysRevC:2008}
W.~H. Lippincott {\em et~al.}
\newblock Scintillation time dependence and pulse shape discrimination in
  liquid argon.
\newblock {\em Phys. Rev. C}, 78(3), 9 2008.

\bibitem{Back:NuclearInstrumentsAndMethodsInPhysicsResearch:2008}
H.~O. Back {\em et~al.}
\newblock Pulse-shape discrimination with the counting test facility.
\newblock {\em Nuclear Instruments and Methods in Physics Research A},
  584:98--113, 1 2008.

\bibitem{Walkowiak:NuclearInstrumentsAndMethodsInPhysicsResearch:2000}
W.~Walkowiak.
\newblock Drift velocity of free electrons in liquid argon.
\newblock {\em Nuclear Instruments and Methods in Physics Research Section A:
  Accelerators, Spectrometers, Detectors and Associated Equipment},
  449(1-2):288--294, 7 2000.

\bibitem{Cao:PhysicalReviewD:2015}
H.~Cao {\em et~al.}
\newblock Measurement of scintillation and ionization yield and scintillation
  pulse shape from nuclear recoils in liquid argon.
\newblock {\em Physical Review D}, 91(9):92007, 5 2015.

\bibitem{Monreal:EprintArxiv14100076:2014}
B.~Monreal.
\newblock Underground physics without underground labs: large detectors in
  solution-mined salt caverns.
\newblock {\em eprint arXiv:1410.0076}, 9 2014.

\bibitem{Biagi:1999nwa}
S.~F. Biagi.
\newblock {Monte Carlo simulation of electron drift and diffusion in counting
  gases under the influence of electric and magnetic fields}.
\newblock {\em Nucl. Instrum. Meth.}, A421(1-2):234--240, 1999.

\bibitem{Collaboration:EprintArxiv150902910:2015}
{LZ Collaboration} {\em et~al.}
\newblock {LUX-ZEPLIN} ({LZ}) conceptual design report.
\newblock {\em eprint arXiv:1509.02910}, 9 2015.

\bibitem{bolotnikov1999studies}
A.~Bolotnikov and B.~Ramsey.
\newblock Studies of light and charge produced by alpha-particles in
  high-pressure xenon.
\newblock {\em Nuclear Instruments and Methods in Physics Research Section A:
  Accelerators, Spectrometers, Detectors and Associated Equipment},
  428(2):391--402, 1999.

\bibitem{Benziger:NuclearInstrumentsAndMethodsInPhysicsResearch:2007}
J.~Benziger {\em et~al.}
\newblock The nylon scintillator containment vessels for the {B}orexino solar
  neutrino experiment.
\newblock {\em Nuclear Instruments and Methods in Physics Research Section A:
  Accelerators, Spectrometers, Detectors and Associated Equipment},
  582(2):509--534, 11 2007.

\bibitem{Bungau:PhysicalReviewLetters:2012}
A.~Bungau {\em et~al.}
\newblock Proposal for an electron antineutrino disappearance search using
  high-rate {Li}$_8$ production and decay.
\newblock {\em Physical Review Letters}, 109(14):141802, 10 2012.

\bibitem{adelmann2014cyclotrons}
A.~Adelmann {\em et~al.}
\newblock Cyclotrons as drivers for precision neutrino measurements.
\newblock {\em Advances in High Energy Physics}, 2014, 2014.

\bibitem{Rott:EprintArxiv151000170:2015}
C.~Rott, S.~In, J.~Kumar, and D.~Yaylali.
\newblock eprint arxiv:1510.00170, 10 2015.

\bibitem{McKinsey:AstroparticlePhysics:2005}
D.~N. McKinsey and K.~J. Coakley.
\newblock Neutrino detection with {CLEAN}.
\newblock {\em Astroparticle Physics}, 22(5-6):355--368, 1 2005.

\bibitem{white2007sign}
J.~T. White, J.~Gao, G.~Salinas, and H.~Wang.
\newblock {SIGN}--a gaseous-neon-based underground physics detector.
\newblock {\em Nuclear Physics B-Proceedings Supplements}, 173:144--147, 2007.

\bibitem{Rielage:PhysicsProcedia:2015}
K.~Rielage {\em et~al.}
\newblock Update on the miniclean dark matter experiment.
\newblock {\em Physics Procedia}, 61:144--152, 2015.

\bibitem{dmtools}
http://dmtools.brown.edu.

\bibitem{Neumeier:Epl:2015_2}
A.~Neumeier {\em et~al.}
\newblock Attenuation of vacuum ultraviolet light in pure and xenon-doped
  liquid argon an approach to an assignment of the near-infrared emission from
  the mixture.
\newblock {\em EPL}, 111(1):12001, 7 2015.

\bibitem{Stenum:TheJournalOfChemicalPhysics:1993}
B.~Stenum, J.~Schou, H.~Sorensen, and P.~G\"{u}rtler.
\newblock Luminescence from pure and doped solid deuterium irradiated by {keV}
  electrons.
\newblock {\em The Journal of Chemical Physics}, 98(1):126, 1993.

\bibitem{Pansky:NuclearInstrumentsAndMethodsInPhysicsResearch:1995}
A.~Pansky {\em et~al.}
\newblock The scintillation of {CF}$_4$ and its relevance to detection science.
\newblock {\em Nuclear Instruments and Methods in Physics Research Section A:
  Accelerators, Spectrometers, Detectors and Associated Equipment},
  354(2-3):262--269, 1 1995.

\bibitem{Jones:2013mfa}
B.~J.~P. Jones {\em et~al.}
\newblock {The Effects of Dissolved Methane upon Liquid Argon Scintillation
  Light}.
\newblock {\em JINST}, 8:P12015, 2013, 1308.3658.

\bibitem{Aprile:1987jt}
E.~Aprile, K.~L. Giboni, and C.~Rubbia.
\newblock {DRIFTING ELECTRONS OVER LARGE DISTANCES IN LIQUID ARGON - METHANE
  MIXTURES}.
\newblock {\em Nucl. Instrum. Meth.}, A253:273--277, 1987.

\bibitem{Tanaka:2004br}
H.~Tanaka {\em et~al.}
\newblock {Study of the electric field dependence on electron drift velocity in
  liquid methane using cosmic ray}.
\newblock In {\em {In *Tsukuba 2004, Radiation detectors and their uses*
  33-42}}, pages 33--42, 2004.

\bibitem{mainz_spectral_atlas}
H.~Keller-Rudek, G.~K. Moortgat, R.~Sander, and R.~Sörensen.
\newblock The {MPI}-{M}ainz {UV/VIS} spectral atlas of gaseous molecules of
  atmospheric interest.
\newblock {\em Earth Syst. Sci. Data}, 5:365--373, 2013.

\bibitem{Holroyd:NuclearInstrumentsAndMethodsInPhysicsResearch:1985}
R.~A. Holroyd and D.~F. Anderson.
\newblock The physics and chemistry of room-temperature liquid-filled
  ionization chambers.
\newblock {\em Nuclear Instruments and Methods in Physics Research Section A:
  Accelerators, Spectrometers, Detectors and Associated Equipment},
  236(2):294--299, 5 1985.

\bibitem{dawson_2014}
J.~V. Dawson and D.~Kryn.
\newblock Organic liquid {TPC}s for neutrino physics.
\newblock {\em Journal of Instrumentation}, 9(07):P07002, 2014.

\bibitem{Rountree:ScienceAndTechnologyOfALowEnergySolar:2010}
S.~A. Rountree.
\newblock {\em Science and Technology of a Low-Energy Solar Neutrino
  Spectrometer (LENS) And Development of the MiniLENS Underground Prototype}.
\newblock PhD thesis, Virginia Tech, 2010.

\bibitem{PhysRevD.87.033005}
X.~Qian {\em et~al.}
\newblock Mass hierarchy resolution in reactor anti-neutrino experiments:
  Parameter degeneracies and detector energy response.
\newblock {\em Phys. Rev. D}, 87:033005, Feb 2013.

\bibitem{doi:10.1146/annurev-nucl-102711-095006}
K.~Scholberg.
\newblock Supernova neutrino detection.
\newblock {\em Annual Review of Nuclear and Particle Science}, 62(1):81--103,
  2012.

\bibitem{Learned:2007zz}
J.~G. Learned, S.~T. Dye, and S.~Pakvasa.
\newblock {Hanohano: A deep ocean anti-neutrino detector for unique neutrino
  physics and geophysics studies}.
\newblock In {\em {Neutrino telescopes. Proceedings, 12th International
  Workshop, Venice, Italy, March 6-9, 2007}}, pages 235--269, 2007, 0810.4975.

\bibitem{Rubbia:09081286:2009}
A.~Rubbia.
\newblock Underground neutrino detectors for particle and astroparticle
  science: the giant liquid argon charge imaging experiment ({GLACIER}).
\newblock {\em 0908.1286}, 8 2009.

\end{thebibliography}

\end{document}